\documentclass{article}

\usepackage{amssymb}
\usepackage{pifont}
\usepackage{amsmath}
\usepackage{graphicx}
\usepackage{bm}
\usepackage{PRIMEarxiv}
\usepackage{float}

\usepackage[utf8]{inputenc} 
\usepackage[T1]{fontenc}    
\usepackage{hyperref}       
\usepackage{url}            
\usepackage{booktabs}       
\usepackage{amsfonts}       
\usepackage{nicefrac}       
\usepackage{microtype}      
\usepackage{lipsum}
\usepackage{fancyhdr}       
\usepackage{graphicx}       
\graphicspath{{media/}}     

\title{Speech Bandwidth Expansion Via High Fidelity Generative Adversarial Networks}

\author{
  Mahmoud Salhab \\
  Lebanese American University \\
  Department of Computer Science and Mathematics \\
  Byblos, Lebanon\\
  \texttt{mahmoud.salhab@lau.edu.lb} 
   \And
  Haidar Harmanani \\
  Lebanese American University \\
  Department of Computer Science and Mathematics \\
  Byblos, Lebanon\\
  \texttt{haidar@lau.edu.lb} 
}

\begin{document}
\maketitle

\begin{abstract}
Speech bandwidth expansion is crucial for expanding the frequency range of low-bandwidth speech signals, thereby improving audio quality, clarity and perceptibility in digital applications. Its applications span telephony, compression, text-to-speech synthesis, and speech recognition. This paper presents a novel approach using a high-fidelity generative adversarial network, unlike cascaded systems, our system is trained end-to-end on paired narrowband and wideband speech signals. Our method integrates various bandwidth upsampling ratios into a single unified model specifically designed for speech bandwidth expansion applications. Our approach exhibits robust performance across various bandwidth expansion factors, including those not encountered during training, demonstrating zero-shot capability. To the best of our knowledge, this is the first work to showcase this capability. The experimental results demonstrate that our method outperforms previous end-to-end approaches, as well as interpolation and traditional techniques, showcasing its effectiveness in practical speech enhancement applications.

\end{abstract}


\section{Introduction}
Speech Bandwidth Expansion (BWE) involves the conversion of a narrowband speech signal into a wideband one. This process is also referred to as Audio Super Resolution, where the objective is to generate a high-resolution speech signal from a low-resolution input containing only a fraction of the original samples \cite{kim2019bandwidthextensionrawaudio}. BWE serves to enhance the quality and perceptibility of narrowband speech, a capability particularly valuable in contexts such as Public Switching Telephone Network (PSTN) environments \cite{litteral1993pstn}. Bandwidth expansion plays a vital role in many systems, and it has shown how the performance of automatic speech recognition systems degrades when encountering a narrowband speech signal \cite{li15d_interspeech}.

Although the critical need for speech transmission bandwidth has diminished in modern times, numerous devices and equipment still operate with, receive, and even store narrowband speech. For example, many Bluetooth headphones continue to function based on narrowband speech \cite{haartsen2000bluetooth}. 

A speech signal can be described as a continuous function, denoted as $f(t)$ over the interval $[0, T]$, where $T$ is duration of the speech in seconds, and $f(t)$ is the amplitude at time $t$. To convert this continuous signal into a discrete form, a sampling technique is employed, in such a way, a value of the signal is captured every $T_{s}$ seconds. This sampling process establishes the sampling rate of the signal, denoted as $F_s$ (in Hz), which can be calculated as the reciprocal of the sampling interval $T_{s}$. Consequently, the continuous function $f(t)$ can be discretized into a vector $x$ comprising samples taken at regular interval $T_{s}$, such that $x = \{ f(T_s), f(2T_s), f(3T_s),\hdots f(nT_s) \} $. In terms of the sample rate, the discretized signal can be represented as $x = \{f(\dfrac{1}{F_s}), f(\dfrac{2}{F_s}), f(\dfrac{3}{F_s}),\hdots f(\dfrac{n}{F_s})\}$. Where $n$ is the number of samples, which can be calculated by $n = \lfloor \dfrac{T}{T_s} \rfloor$. The sampling rate of a signal can vary significantly, spanning from 4 KHz, typically associated with low-quality telephone speech, to 48 KHz, indicative of high-quality speech/music. 

According to the Nyquist–Shannon sampling theorem proposed by Shannon \cite{Shannon1949}, when a signal is sampled at a rate of $F_s$ Hz, the maximum bandwidth $B$ of that signal that guarantees no aliasing is $B = \dfrac{F_s}{2}$. Thus, to expand the bandwidth $B$ by a factor of $\mathbf{s}$, the sampling rate must also be scaled by $\mathbf{s}$. Formally, in the context of a narrowband speech signal with a sample rate and bandwidth $F_{low}$ and $B_{low}$ respectively, bandwidth expansion entails scaling the sampling rate and bandwidth by an upsampling ratio $\mathbf{s}$. This results in a wideband speech signal with $F_{high}$  = $\mathbf{s} \times F_{low}$, $B_{high}$  = $\mathbf{s} \times B_{low}$, and $|x_{high}| = \mathbf{s} \times |x_{low}|$, where $x_{high}$ represents the wideband speech signal and $x_{low}$ represents the narrowband one. Consequently, the bandwidth expansion or the audio super-resolution problem is akin to reconstructing the missing frequency content between $B_{high}$ and $B_{low}$.

In this study, we propose utilizing a high-fidelity generative adversarial network for speech bandwidth expansion across different upsampling ratios. Unlike traditional cascaded techniques, our method is end-to-end, requiring only a single model for both training and inference. To further enhance performance, we introduce a unified model capable of processing various upsampling ratios, eliminating the need for separate models for each upsampling ratio. Additionally, we evaluate our model in zero-shot settings, where the model processes previously unseen narrowband speech signals to scale them to unseen upsampling ratio. To the best of our knowledge, this is the first work to introduce zero-shot capability for neural speech bandwidth expansion.

This paper is organized as follows. Section \ref{sec:related} presents a literature review, establishing the context and background for our research, followed by our methodology in section \ref{sec:methodology}. Sections \ref{sec:experiments} and \ref{sec:results} present the experiments and the results, and lastly, we conclude in section \ref{sec:conc} with a summary of potential future work.

\section{Related Work}
\label{sec:related}

Bandwidth expansion, also referred to as audio super-resolution in the literature, and it has been a subject of study for decades due to its significance in many systems \cite{781522,Vaseghiinproceedings1,6349783,a0d4fcfbd7c44c8b8395ed30c8ef1337, Ekstrand2002, Larsen2005, Jax2003, Qian2002, Nour-Eldin2008}. Early studies focused on estimating the spectral envelope of the high-frequency band and using excitation generated from the low-frequency band to recover the high-frequency spectrum \cite{Iser2008}. Traditional techniques such as Gaussian mixture models, linear predictive coding, and hidden Markov models have also been used \cite{7953218inproceedings, Bradbury2000,6495700}. However, these methods generally perform worse compared to neural networks \cite{8063328}. Additionally, matrix factorization techniques trained on very small datasets have been proposed to mitigate the computational cost of factorizing matrices \cite{Liang2013, Bansal2005}.

Recent advancements in end-to-end deep neural networks generate wideband signals directly from narrowband signals without the need for feature engineering. For instance, \cite{7178801} proposed training a deep neural network as a mapping function, using log-spectrum power as the input and output features to perform the required nonlinear transformation. A dense neural network with three hidden layers of size 2048 and ReLU activation function proposed in \cite{li15d_interspeech}, showed that this method is preferred over Gaussian mixture models in 84\% of cases in a user study they made.

Inspired by image super-resolution algorithms, which use machine learning techniques to interpolate a low-resolution image into a higher-resolution one, a convolutional U-net contains successive downsampling and upsampling blocks with skip connections proposed in \cite{kuleshov2017audiosuperresolutionusing}. Building on that,  a neural network component named Temporal Feature-Wise Linear Modulation (TFiLM) introduced in \cite{birnbaum2021temporalfilmcapturinglongrange}. TFiLM captures long-range input dependencies in sequential inputs by combining elements of convolutional and recurrent approaches in a U-net-like architecture. Furthermore, it modulates the activations of a convolutional model using long-range information captured by a recurrent neural network. A block-online variant of the temporal feature-wise linear modulation (TFiLM) model to achieve bandwidth extension was proposed by \cite{Nguyen_2022}. This architecture simplifies the UNet backbone of the TFiLM to reduce inference time and employs an efficient transformer at the bottleneck to alleviate performance degradation. It also utilizes self-supervised pretraining and data augmentation to enhance the quality of bandwidth-extended signals and reduce sensitivity with respect to downsampling methods.

Attention-based Feature-Wise Linear Modulation (AFiLM) \cite{rakotonirina2021selfattentionaudiosuperresolution} proposed a network with a U-net-like architecture for audio super-resolution that combines convolution and self-attention. AFiLM uses a self-attention mechanism instead of recurrent neural networks to modulate the activations of the convolutional model.

In \cite{8706637}, to model the distribution of the target high-resolution signal conditioned on the log-scale mel-spectrogram of the low-resolution signal researchers utilized the WaveNet \cite{oord2016wavenetgenerativemodelraw} model. Furthermore, \cite{macartney2018improvedspeechenhancementwaveunet} studied the use of the Wave-U-Net architecture for speech enhancement.

Generative Adversarial Networks (GANs) \cite{goodfellow2014generativeadversarialnetworks} are utilized in \cite{kumar2020nuganhighresolutionneural}, where an additional third component is proposed in the TTS pipeline for neural upsampling. This converts lower resolution audio (16-24 kHz) to full high-resolution audio (44.1 kHz). Additionally, researchers explored the use of a diffusion probabilistic model for audio super-resolution, which is designed based on neural vocoders \cite{Lee2021}.

A Glow-based waveform generative model for performing audio super-resolution was proposed in \cite{zhang2021wsrglowglowbasedwaveformgenerative}. Specifically, the integration of WaveNet and Glow directly maximizes the exact likelihood of the target high-resolution (HR) audio conditioned on low-resolution (LR) information. To exploit audio information from low-resolution audio, an LR audio encoder and an STFT encoder are proposed, which encode the LR information from the time domain and frequency domain, respectively.

A neural vocoder-based speech super-resolution method (NVSR) was proposed by \cite{Liu2022}, capable of handling various input resolutions and upsampling ratios. NVSR follows a cascaded system structure, comprising a mel-bandwidth extension module, a neural vocoder module, and a post-processing module.

A diffusion-based generative model designed to perform robust audio super-resolution across a wide range of audio types named AudioSR proposed in \cite{liu2023audiosrversatileaudiosuperresolution}. AudioSR designed specifically to enhance the quality of sound effects, music, and speech.

\section{Methodology}
\label{sec:methodology}

Given a dataset $ \mathcal{D} = \{(x_{1}, \hat{x}_{1}), (x_{2}, \hat{x}_{2}), \ldots, (x_{M}, \hat{x}_{M})\} $, where each pair consists of the same speech signal sampled at different frequencies, we define $\hat{x}_{m}$ as the signal sampled at a lower frequency $F_{low}$ with bandwidth $B_{low}$, and $x_{m}$ as the signal sampled at a higher frequency $F_{high}$ with bandwidth $B_{high}$. Both signals originate from the same source. Mathematically, $F_{high} = \mathbf{s} \times F_{low}$ and $B_{high} = \mathbf{s} \times B_{low}$, such that $\mathbf{s} > 1$, where $\mathbf{s}$ is the super-resolution/upsampling ratio. Formally, we represent $x_{m}$ as $x_{m} = \{ x_{1}, x_{2}, \ldots, x_{N} \}$ and $\hat{x}_{m}$ as $\hat{x}_{m} = \{ \hat{x}_{1}, \hat{x}_{2}, \ldots, \hat{x}_{ \lfloor \frac{N}{\mathbf{s}}\rfloor}  \}$.

Our goal is to expand the bandwidth of the given signal $\hat{x}$ by a factor of $\mathbf{s}$, resulting in a speech signal with an increased sampling frequency and bandwidth. The objective of neural super-resolution is to construct a function $\mathcal{F}_{\theta}$ such that $\acute{x} = \mathcal{F}_{\theta}(\hat{x})$, where $\acute{x}$ is the upsampled or reconstructed version of the input speech signal $\hat{x}$ with bandwidth $B_{high}$. In such a way, $\acute{x}$ is as close as possible to the ground truth one $x$. The aim is to develop an effective yet efficient function $\mathcal{F}_{\theta}$. A straightforward approach involves designing $\mathcal{F}_{\theta}$ as a simple DNN \cite{Rumelhart1986LearningRB} or CNN \cite{6795724} that minimizes the following objective function:

\begin{equation}\label{eq:1}
\min_{\theta} \sum_{m=1}^{M}\sum_{n=1}^{N}(\mathcal{F}_{\theta}(\hat{x}_{m})_{n} - x_{m,n})^2
\end{equation}

If the network is not properly designed, several issues may arise. Speech signals tend to be lengthy, making it computationally inefficient to train a complex model. Additionally, distance-based objective functions have inherent limitations. Training neural networks with such objectives often results in blurry outputs in computer vision, known as the "softness" issue \cite{zhang2016colorfulimagecolorization, pathak2016contextencodersfeaturelearning}, and similarly in speech. Therefore, our objective is to build an efficient $\mathcal{F}_{\theta}$ and use an appropriate objective function to train $\mathcal{F}_{\theta}$ on the dataset, ensuring the reconstructed speech signal closely matches the high-resolution signal. To achieve this, we employed an adversarial loss to eliminate the softness issue and utilized a convolutional model architecture, which will be elaborated upon in the following sections.

\subsection{Model}

\begin{figure}
\centering
\includegraphics[scale=0.45]{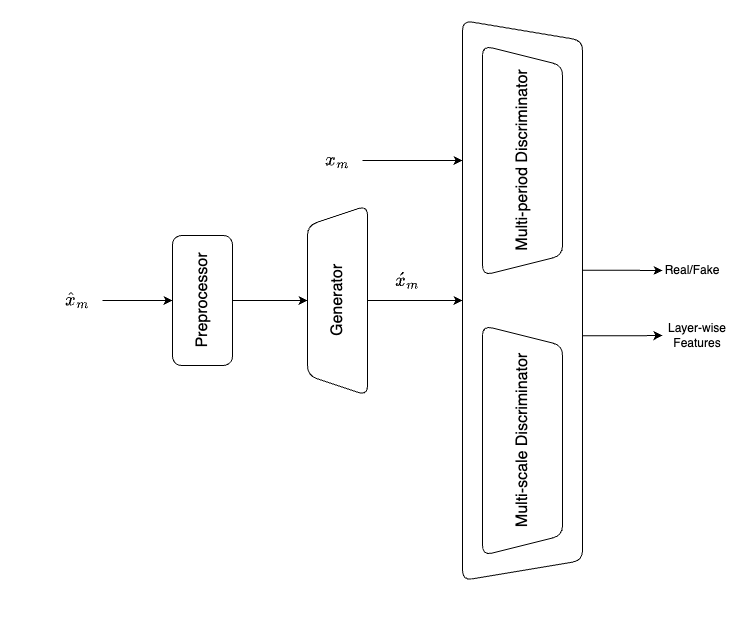}
\caption{Complete architecture of the model.}
\label{fig:model-arch}
\end{figure}

In the realm of speech audio, where signals exhibit sinusoidal characteristics with varying periods, accurately representing these periodic patterns is paramount for synthesizing authentic, high-fidelity speech from an original, low-fidelity source. Building upon the methodology in \cite{hifi-gan}, our model comprises a generator and two distinct discriminator types: multi-scale and multi-period discriminators. These elements undergo adversarial training, supplemented by two auxiliary loss functions to boost training stability. The multi-period discriminator comprises several sub-discriminators, each targeting specific periodic segments within the raw waveforms. Complementing this, the generator module incorporates multiple residual blocks, each adept at processing patterns of different lengths concurrently. This design ensures the model comprehensively captures the diverse range of periodic characteristics present in speech audio.
The entire model architecture is illustrated in Figure \ref{fig:model-arch}, and detailed explanations of each subcomponent follow in the subsequent sections.

\subsubsection{Generator}
The generator is a convolutional neural network that's entirely dedicated to upscaling. It begins with a mel-spectrogram input generated within the input speech signal's preprocessor module, and uses transposed convolutions to gradually match the temporal resolution of the high-resolution speech signal $x_{m}$. Each transposed convolution operation is accompanied by a multi-receptive field fusion (MRF) module. This module is designed to analyze patterns of different lengths simultaneously. It achieves this by summing the outputs from several residual blocks. These blocks employ various kernel sizes and dilation rates to create a wide range of receptive field patterns. The architecture of the generator is identical to the generator proposed in \cite{hifi-gan}.

\subsubsection{Discriminator}

Following the methodology presented in \cite{hifi-gan}, the design of the discriminator tackles two critical challenges. Firstly, it must adeptly capture the extended dependencies within the input speech signal. Secondly, given the sinusoidal nature of the input with variable periods, it needs to discern the diverse periodic patterns inherent in the audio data. Thus, we adopt a similar strategy as detailed in the original work \cite{hifi-gan}. This involves implementing a multi-period discriminator (MPD) composed of multiple sub-discriminators, each responsible for analyzing a segment of the periodic signals within the input audio. Additionally, a multi-scale discriminator (MSD) is employed to detect sequential patterns and long-term dependencies effectively.

\subsection{Training Loss}
For the generator and discriminator, the training objectives follow the approach proposed in \cite{mao2017squares}, where the binary cross-entropy loss function is replaced with the least square error loss. This substitution helps prevent the vanishing gradient issue. Additionally, an extra objective function is included, specifically mel-spectrogram reconstruction as proposed by \cite{isola2018imagetoimage}, along with discriminator feature-wise matching across layers as in \cite{kumar2019melgan}. The discriminator is trained to classify ground truth samples as 1 and samples synthesized by the generator as 0. The generator, on the other hand, is trained to deceive the discriminator by enhancing the sample quality so that it is classified as close to 1 as possible. The objective function for the discriminator $D_{\phi}$ shown in the Equation \ref{eq:disc-obj} and for the generator $G_{\theta}$ is expressed in the Equation \ref{eq:gen-obj-1}.

\begin{equation}\label{eq:disc-obj}
\min_{D_{\phi}} \mathbb{E} [(1 - D_{\phi}(x))^{2} + D_{\phi}(G_{\theta}(\hat{x}))^{2}]
\end{equation}

\begin{equation}\label{eq:gen-obj-1}
\min_{G_{\theta}} \lambda_{adv} \mathcal{J}_{adv} + \lambda_{mel} \mathcal{J}_{mel} + \lambda_{feat} \mathcal{J}_{feat}
\end{equation}

\begin{equation}\label{eq:gen-obj-2}
\mathcal{J}_{adv}  = \mathbb{E}[(1 - D_{\phi}(G_{\theta}(\hat{x})))^{2}] 
\end{equation}

\begin{equation}\label{eq:gen-obj-3}
\mathcal{J}_{mel} = \mathbb{E}[ \Vert \psi(x) - \psi(G_{\theta}(\hat{x})) \Vert_{1} ]
\end{equation}

\begin{equation}\label{eq:gen-obj-4}
\mathcal{J}_{feat} = \mathbb{E} [ \sum_{k=1}^{K} \frac{1}{N_{k}}  \Vert D_{\phi}^{k}(x) - D_{\phi}^{k}(G_{\theta}(\hat{x})) \Vert_{1}]
\end{equation}

In these equations, $\lambda_{adv}$, $\lambda_{mel}$, and $\lambda_{feat}$ are hyperparameters that balance the contributions of the adversarial, mel-spectrogram reconstruction, and feature matching term, respectively. $\psi$ represents the function that computes the mel-spectrogram, $D_{\phi}^{k}$ denotes the output of the $k^{th}$ layer of the discriminator, and $N_{k}$ is the number of elements in the $k^{th}$ layer.

\section{Experiments}
\label{sec:experiments}

\subsection{Dataset}

Similar to prior studies, we utilized the VCTK corpus \cite{Yamagishi2019CSTRVC} for both training and evaluation purposes. The VCTK dataset comprises approximately 44 hours of audio recordings, from 109 distinct speakers, offering a variety of voices and accents including Scottish, Indian, and Irish, among others. The dataset originally uses a bit width of 16-bit PCM with a sample rate of 48 kHz. Therefore, we first resampled the entire dataset to a sample rate of 16 kHz while maintaining the same bit width. To create low-resolution audio, we applied an order 8 Chebyshev type I low-pass filter to the original 16 KHz signals, followed by subsampling according to the desired upsampling ratio.

Our research centers on a multi-speaker task, training on the first 100 VCTK speakers and testing on the remaining speakers \footnotemark, in alignment with the methodology presented in \cite{kuleshov2017audiosuperresolutionusing}. \footnotetext{The speaker IDs used for testing are p347, p351, p360, p361, p362, p363, p364, p374, p376.}

\subsection{Evaluation Metric}

To evaluate the performance of our proposed method and to measure the quality of generated audio samples by comparing them to the actual high-resolution audio, and following previous works, we assess the reconstruction quality of individual frequencies using the Log Spectral Distance (LSD). The LSD is calculated as shown in Equation \ref{eq:lsd}.

\begin{equation}\label{eq:lsd}
LSD(x, \acute{x}) = \frac{1}{T} \sum_{t=1}^{T} \sqrt{\frac{1}{F} \sum_{f=1}^{F} (\mathcal{P}(x)_{t,f} - \mathcal{P}(\acute{x})_{t,f})^{2}}
\end{equation}

\begin{equation}\label{eq:lsd-p}
\mathcal{P}(z) = log_{10} |STFT(z)|^{2}
\end{equation}

Here, $x$ represents the reference wideband speech signal, and $\acute{x}$ is the reconstructed speech signal from the model. $\mathcal{P}(x)$ denotes the log-spectral power magnitudes calculated using Equation \ref{eq:lsd-p}, where $STFT$ refers to the Short-Time Fourier Transform. Variables $f$ and $t$ correspond to the frequency and frame index, respectively. A lower LSD values indicate better frequency reconstruction. Following \cite{kuleshov2017audiosuperresolutionusing} we used frames of length 2048 to calculate the STFT in our experiments.

\subsection{Experimental Setup}
For all the experiments, we used different upsampling ratios $\mathbf{s}$, specifically 2, 4, and 8. For each upsampling ratio, we trained a separate model, each for 500K steps. Our models were trained on a single machine with two NVIDIA 3080 TI GPUs, using a global batch size of 16 and the same model configuration as version 1 described in \cite{hifi-gan}.

Since the preprocessing handles the upsampling, we also trained a unified model on all the aforementioned upsampling ratios. Additionally, we investigated the ability of the proposed technique to work in zero-shot settings, where the model is given a signal and tasked with generating a signal with a different upsampling ratio than those it was trained on. We examined whether the model could generalize and predict the signal. The unified model was trained similarly to the single upsampling ratio models, but for longer, specifically for 1.5M steps.

All the models were trained on mel-spectrogram input, calculated using $80$ mel-filter bank with an FFT size of $1024$, a window size of $1024$, and a hop length of $256$. For the generator loss terms, we set $\lambda_{adv} = 1.1$, $\lambda_{mel} = 50$, and $\lambda_{feat} = 2$. In addition to that, the AdamW optimizer was used with $\beta_{1} = 0.8$ and $\beta_{2} = 0.999$, with a weight decay $\lambda = 0.01$, and the learning rate used on each epoch calculated using Equation \ref{eq:lr}, where $lr_{init}$ is the initial learning rate set to $1.5 \times 10^{-4}$, $epoch$ is the current training epoch, and $\gamma$ is the learning rate decaying factor set to $0.999$.

\begin{equation}\label{eq:lr}
lr(epoch) = \gamma^{epoch} * lr_{init} 
\end{equation}

\section{Results}
\label{sec:results}

The results of our experiments are presented in Table \ref{tab:results}, which displays the Log Spectral Distance (in dB) for various upsampling ratios, specifically 2, 4, and 8. The table compares the performance of different baselines with two scenarios: first, when each upsampling ratio is trained on a standalone model, labeled as "Single," and second, when all upsampling ratios are trained jointly using a single model, labeled as "Unified." Our model consistently outperforms all end-to-end baselines. When compared to cascaded models such as NVSR \cite{Liu2022}, our approach performs better for high upsampling ratio (i.e., $\mathbf{s}=8$) and is comparable for $\mathbf{s}=4$, although cascaded models outperform our model for low upsampling ratio (i.e., $\mathbf{s}=2$).

\begin{table}[ht]
\centering
\caption{Analysis of Speech Bandwidth Expansion at Upsampling Ratios of 2, 4, and 8.}

\label{tab:results}
\begin{tabular}{lcccc}
\toprule
\toprule
\textbf{Approach} & \textbf{s = 2} & \textbf{s = 4} & \textbf{s = 8} & \textbf{End-to-End} \\
\midrule
\midrule
Baseline & 3.464 & 5.17 & 6.08 & N/A \\
AudioUNet \cite{kuleshov2017audiosuperresolutionusing} & 3.1 & 3.5 & N/A & \checkmark \\
TFNet \cite{Lim2018TimeFrequencyNF} & N/A & 1.27 & 1.9 & \checkmark \\  
Temporal FiLM \cite{birnbaum2021temporalfilmcapturinglongrange} & 1.8 & 2.7 & 2.9 & \checkmark \\  
MU-GAN \cite{kim2019bandwidthextensionrawaudio} & 2.14 & 2.72 & N/A & \checkmark \\ 
AFiLM \cite{rakotonirina2021selfattentionaudiosuperresolution} & 1.7 & 2.3 & 2.7 & \checkmark \\  
NVSR \cite{Liu2022} & 0.78 & 0.95 & 1.07 & \ding{55} \\ 
\midrule

Ours (Single) & 0.9 & 0.98 & 1.047 & \checkmark \\
Ours (Unified) & 0.923 & 0.98 & 1.0 & \checkmark \\
\bottomrule
\bottomrule
\end{tabular}
\end{table}

\begin{figure}
\centering
\includegraphics[scale=0.31]{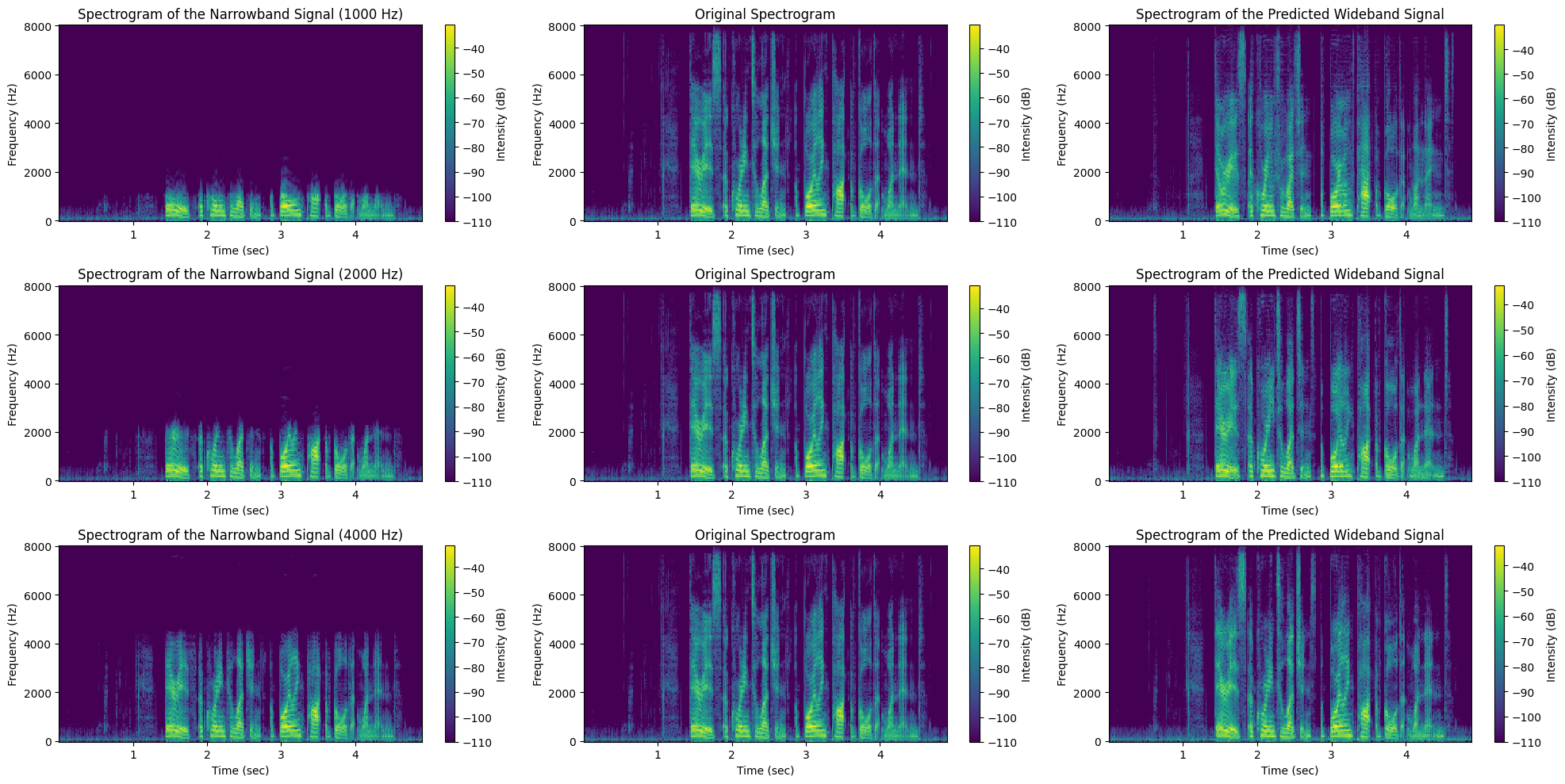}
\caption{Spectrogram Analysis of Narrowband to Wideband Speech Reconstruction with Varying Upsampling Ratios ($\mathbf{s}=8$, $\mathbf{s}=4$, $\mathbf{s}=2$)}
\label{fig:spec}
\end{figure}

A test sample is shown in Figure \ref{fig:spec}, including the input narrowband speech signal, the reconstructed wideband speech signal, and the original wideband speech signal. The first column shows the input narrowband speech signal, the middle column displays the target wideband speech signal, and the last column presents the predicted wideband speech signal using our trained model. The first row corresponds to an upsampling ratio of $\mathbf{s}=8$, the second row to $\mathbf{s}=4$, and the third row to $\mathbf{s}=2$. It is evident that our approach effectively addresses the issue of oversmoothness, successfully reconstructing the missing frequencies.

\begin{figure}[H]
\centering
\includegraphics[scale=0.85]{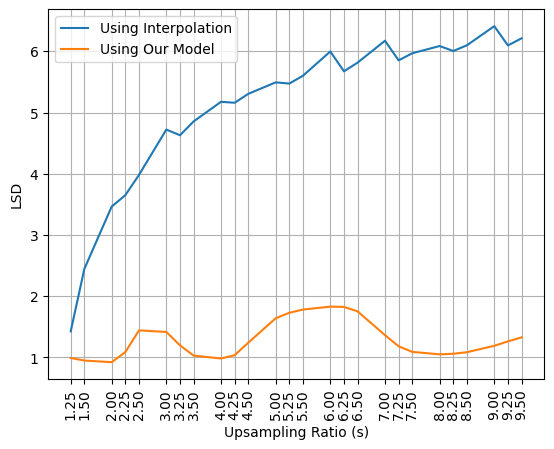}
\caption{Performance comparison of our unified model across various upsampling ratios, demonstrating its ability to handle unseen upsampling ratios with maintained low Log Spectral Distance (LSD) compared to traditional interpolation methods}
\label{fig:unseen-scalign-factors}
\end{figure}

The performance of our unified model across various upsampling ratios on the test set, including those it was trained on (i.e., 2, 4, 8), is illustrated in Figure \ref{fig:unseen-scalign-factors}. Compared to the baseline Fast Fourier Transform (FFT)-based interpolation technique, our method demonstrates the capability to handle even unseen upsampling ratios. Notably, the trained upsampling ratios exhibit the lowest Log Spectral Distance (LSD). Additionally, it is evident that traditional interpolation-based techniques for bandwidth expansion result in increased LSD as the upsampling ratios rises (i.e decreasing the input bandwidth signal). Conversely, our method maintains a bounded LSD, even for upsampling ratios not encountered during training which shows that our model can run in zero-shot settings.

\section{Conclusion}
\label{sec:conc}
In this work, we presented an end-to-end approach for tackling speech bandwidth expansion using a high-fidelity generative adversarial network trained across different super-resolution/upsampling ratios. Empirically, our method surpasses various end-to-end baselines and achieves comparable results when compared with cascaded approaches. Additionally, we demonstrated the scalability of our method to unseen upsampling ratios during training in zeros-shot setting. These findings underscore the potential of our unified model architecture, which not only simplifies the training and deployment process but also enhances performance across varying levels of upsampling ratios. Moving forward, our work opens avenues for further exploration and application of neural speech bandwidth expansion in real-world scenarios.

\bibliographystyle{unsrt}  
\bibliography{references}

\end{document}